# Vibronic Landscape of Excitons in Photosynthetic Antenna

Manuel J. Llansola-Portoles[1*], James Sturgis[1,2*], Andrew Gall[1*], Andrew Pascal[1], Leonas Valkunas[3], Bruno Robert[1#]

[1] : Institute of Integrative Biology of the Cell, CEA, CNRS, Université Paris Saclay, CEA Saclay 91191 Gif sur Yvette Cedex, France

[2]: present address : Laboratoire d'ingénierie des systèmes moléculaires, CNRS, Aix-Marseille University, 31 Chemin Joseph Aiguier, Marseille Cedex 9 13402, France

[3] : Department of Molecular Compound Physics, Centre for Physical Sciences and Technology, Sauletekio Avenue 3, LT-10257 Vilnius, Lithuania

[*]:these authors contributed equally to this work

[#]: to whom correspondence should be addressed

**Light-harvesting and excitation energy transfer in photosynthesis generally involve chlorophyll-molecules, maintained by their host proteins at short distances from each other, this resulting in excitonic coupling. The transfer of excitation energy to the reaction centers consists of exciton migration and relaxation within and between photosynthetic proteins. The dynamics of this process depends on the vibrational modes resonant with the energy gaps between the participating excited states. The precise structure and vibrational landscape of excitons is thus essential knowledge to understand the amazing efficiency of photosynthesis. In this work, we characterize the vibrational properties of excitons in light-harvesting proteins from purple photosynthetic bacteria, which remarkably unveil on how many bacteriochlorophylls they reside and in which proportions. Vibrational spectra obtained from bacteriochlorophylls in proteins generally contain additional vibronic contributions when compared to isolated pigments, opening additional pathways for vibrationally-assisted excitation energy transfer. In contrast, the absence of new vibronic contributions in the spectra of chlorophyll -containing photosynthetic proteins above 100 $cm^{-1}$ suggests that in oxygenic photosynthesis, vibrationally-assisted excitation energy transfers occurs through vibrational modes of chlorophyll molecules in equilibrium configuration.**

The large redshift of the electronic absorption transition of bacteriochlorophyll (BChl) a molecules observed in the light harvesting proteins (LH) from purple bacteria, suggested, as early as in the 70's, that these molecules were in strong excitonic interaction [1], the intensity of which was estimated between 200 and 440 cm$^{-1}$ [ref 2-4]. As these molecules are likely to possess similar site energies, the excited electronic state created by photon absorption in such systems must reside on an excitonically-coupled cluster of BChl molecules. The presence of delocalized excitons is now widely accepted in purple bacteria, and it is generally recognized that they confer to photosynthetic proteins many of their electronic and functional properties[4]. In photosynthetic proteins from plants and algae, the situation more diverse. Although several examples of chlorophyll (Chl) a molecules displaying red-shifted absorption exist, for instance in the Light-Harvesting Complexes (LHC) from photosystem 1, the large majority of Chls absorb at positions close to those observed in solvents, and the excitonic coupling between Chls in these proteins is much weaker, estimated in the tens of cm$^{-1}$ [ref 3]. However, even with weak excitonic coupling, considering the Chl molecules as isolated in their excited state does not allow to properly model either the electronic properties of the LHC, or the yields and rates of the excitation energy transfers cascade occurring photon absorption. In this case again, an excitonic picture of the electronic properties is absolutely required to build-up a satisfactory model. Indeed, advanced modelling of the cascade of events following photon absorption in these proteins involve excitons, and can give quite precise propositions for the exciton structure (see *e.g.* [5]). Many parameters enter into play to determine excitons properties [4]. While the strength of excitonic coupling favours excitation delocalization, the individual electronic properties of the molecules in interaction, or their site energies, to which we have most often no experimental access, also influence the exciton structure. The more similar these are, the more evenly distributed the excitation in a particular exciton state will be. Energy gaps between the individual excited states involved will, on the contrary, tend to localize the exciton on the moiety with the lower energy excited state. Unfortunately, there is still no direct experimental description of the actual molecular structure of photosynthetic excitons, which is frustrating, considering their importance in the early steps of photosynthesis.

Once the exciton is created, rapid reorganization, including interactions with the surrounding dynamics (both local molecular vibrations and the phonon bath)[4] will govern its relaxation. In all light-harvesting proteins, exciton relaxations and transfers leading to the lowest energy excited state occur in less than a picosecond, while excitation energy transfers between complexes are slower (few picoseconds)[6]. Excitation energy migration thus largely involve relaxed excitons, and these are precisely what we studied here. Characterization of the electronic structure of these excited delocalized electronic states is a prerequisite for developing a fundamental understanding their photochemistry. The specific vibrational modes coupled with the electronic transition rule vibrationally-assisted excitation energy transfers and are necessary for understanding energy migration in photosynthesis.

Light harvesting (LH) proteins from purple photosynthetic bacteria, are generally composed of circular arrays of short, transmembrane peptides dimers [7,8]. Each of these peptides binds a Bchl molecule in the membrane core, which results in the formation of a circular array of strongly excitonically coupled pigments. In the core LH (or LH1), this ring is large enough to surround, totally or partially the reaction center. The LH1 protein from *Blastochloris* (*Blc*) *viridis* possesses a structure slightly different from the "typical" LH1. It contains an additional peptide, which does not bind a Bchl, probably stabilizing the protein structure and it is constituted of a 17-mer ring of peptide trimers around the reaction center, binding 34 BChl *b* [9]. The electronic absorption transition of *Blc viridis* LH1 is observed at 1008 nm at room temperature, redshifted by *ca* 2300 cm$^{-1}$ as compared to that of the isolated BChl b pigment (observed at *ca* 820 nm in solvents). Upon lowering the temperature to 4K, this transition shifts even more to the red, at 1049 nm (figure 1) [10]. At this temperature, it displays clear heterogeneity [11], and was proposed to be composed of at least three absorption bands [12].

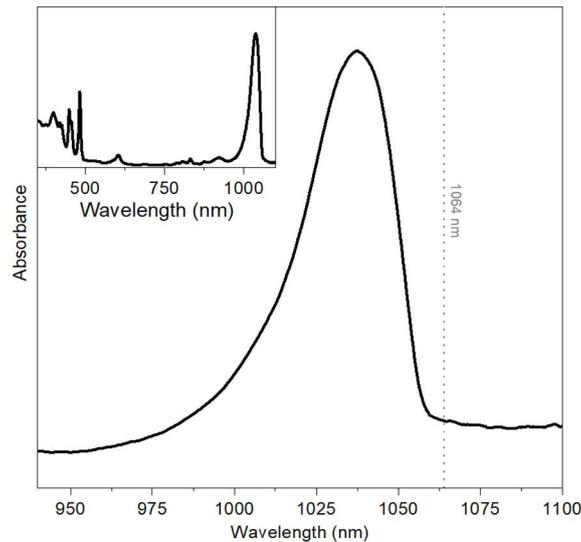

**Figure 1 |** Blc viridis RC-LH1 absorption in the Qy electronic transition region at 4K. Inset: Absorption spectrum (full range) of these RC-LH1. The dotted grey line marks the position of the excitation used to obtain FLN spectra from these proteins.

Fluorescence Line Narrowing (FLN) spectroscopy is a cryogenic technique that can be used to obtain high-resolution fluorescence spectra, containing selective information on those vibrational modes coupled with the electronic transition of the fluorophore, including modes coupled to relaxed excitons in photosynthetic pigments [13]. In order to obtain FLN spectra in excitonically coupled systems of pigments such like LH proteins, it is necessary to excite directly the lowest energy excited state in the system, in order to avoid blurring the resolution by electronic or vibrational energy transfers. Fluorescence polarization experiments, performed in membranes, *i.e.* on unperturbed RC-LH1 from *Blc viridis*, showed that full depolarization could be observed only when the excitation reaches 1050 nm [10]. At this wavelength, polarization abruptly increases to 0.5, indicating that no excitation energy transfer occurs after the absorption of the photon (in a circular array, energy transfer directly perturbs the fluorescence polarization). We thus performed FLN experiments using the 1064 nm excitation beam of a continuous Nd:YAG laser, using a RC-LH1 membrane preparations to avoid eventual perturbation on the structure of the BChl b ring.

As shown in extended data figure 1, clear fluorescence line narrowing is observed upon temperature lowering when exciting *Blc viridis* RC-LH1 complexes at this wavelength, revealing many vibrational contributions. Indeed in these complexes there are many more vibrational contributions than in FLN spectra from isolated BChl b (see as an example the mid-frequency region, figure 2 and extended data figure 2).

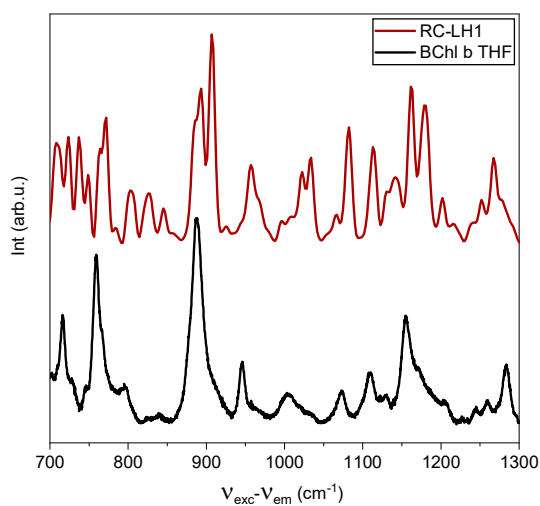

**Figure 2|** Fluorescence Line Narrowing spectra at 4 K (mid frequency region) of RC-LH1 from *Blc. viridis* (red line, excitation at 1064nm) and isolated BChl b (black line, excitation at 822nm).

In these spectra, the intensity of the bands directly correlates with the coupling of the vibrational modes to the excited electronic transition. The electronic structure of the LH1 exciton and its vibrational environment confers intensity to bands which are barely visible in isolated BChl b spectra For example, the mode at 1615 cm$^{-1}$ is intense in LH1 spectra, while its counterpart, expected at 1590 cm$^{-1}$ is barely visible in spectra of isolated pigment (see extended data figure 2). This makes attribution and comparison of these spectra difficult. Nevertheless, in several cases, the additional contributions observed in LH1 spectra arise from splitting of single bands present in BChl b spectra to give a doublet (table 1). All of these doublets originate from vibrational modes sensitive BChl conformation (table 1) [14], or from modes whose frequency is sensitive to the nature of the pigment-protein interaction, suggesting dependence on specific configurations of BChl molecules [15].

**Table 1 :** Frequencies in cm$^{-1}$ of the major band splitting observed in FLN spectra of *Blc. viridis* RC-LH1 relative to isolated BChl b in THF.

| RC-LH1 *Blc viridis* | 712/725 | 764/771 | 892/907 | 997/1007 | 1132/1142 | 1160/1180 | 1533/1544 |
|---|---|---|---|---|---|---|---|
| BChl b | 718** | 762** | 892* | 1008* | 1131* | 1158* | 1533* |

* : modes shown to be sensitive to BChl conformation[14]
**: modes reported to shift frequencies in different LH proteins[15]

Superradiance experiments, performed on *Blc viridis* RC-LH1, showed that, at similar temperature, fluorescence occurs from an excited state residing on more than one BChl molecule [10]. The FLN spectra from RC-LH1 of *Blc viridis* actually must contain contributions from at least two inequivalent BChl molecules, more precisely, from one BChl possessing a configuration and vibrational signature close to that observed in solution and from another one, in a more distorted configuration. From resonance Raman, circular dichroism experiments it has already been suggested that the BChl dimer constituting the repeat unit in circular LH should comprise two molecules in different conformation [16,17]. In *Rhodopseudomonas acidophila* LH2, it was further concluded from crystallographic studies, that while the set of BChl molecules bound to the α polypeptide of this complex is in its relaxed configuration, the set of BChl bound to the other polypeptide is in a distorted, saddled configuration[18]. The observed band splittings in FLN spectra of RC-LH1 from *Blc viridis* thus demonstrate that, at 4K, the transition from the excited state occurs to the ground states of two different molecules, and thus that the exciton resides at least on one repeat dimer of the ring. The similar intensity of the components of the split bands suggests that this exciton be equally partitioned between the molecules constituting such a dimer. However, such equal partitioning is somehow unlikely, as BChl molecules with different conformations are expected to exhibit different site energies (this difference was evaluated to 300 $cm^{-1}$ in LH2s) [16], resulting in a concentration of the exciton on the molecule with lower site energy. Superradiance experiments suggested the length of the exciton to be between three and four molecules[10]. In LH2, the length of the exciton was estimated a to be little shorter (2.8 +/- 0.4) [19]. Taking into account these results, which all conclude the exciton length should be superior to two, the simplest interpretation of *Blc viridis* FLN spectra is thus that, at 4K, the exciton resides on three BChl molecules, with a 25/50/25 partition to account for the similar intensity of the contributions of relaxed and strained BChl molecules.

In addition to vibrational band splittings, FLN spectra of RC-LH1 from *Blc viridis* contain two intense bands, either absent or very weak in spectra of isolated BChl b, located at 293 and 482 $cm^{-1}$, the origin of which remains to be determined. FLN spectra of *Blc viridis* contain bands arising only from modes conjugated with the Qy BChl transition. Thus no band can be observed for instance at 658 or 1348 cm$^{-1}$, where the main contributions from modes conjugated with the Qx transition are expected [20]. The coupling of these additional modes with the electronic transition may open additional channels for excitation energy transfer in the photosynthetic apparatus through vibrational assistance.

Bchl-containing antenna include the antenna from purple photosynthetic bacteria, which are all built according to the same scaffold, and also the Fenna Matthews Olson (FMO) protein, a soluble protein binding BChl a from green sulphur bacteria. This protein binds seven BChl a, and displays a redshifted absorption at low temperature [21]. The structure of this transition was debated, hole burning studies showing the presence of three components in it[22] while FLN studies could only identify a single component[23]. It is generally considered that this exciton is mostly concentrated on one BChl a molecule, to which after photon absorption, the excitation energy is transferred[24]. In FLN spectra of this protein[23], no splitting of bands is observed when comparing with spectra of isolated BChl a, which is consistent with the predicted localization of the exciton on a single molecule (see figure 3 for a direct comparison). However once again additional bands can be observed (figure 3), with different frequencies from those observed for the LH1-RC from *Bls viridis* at 539, 811 and 851 cm$^{-1}$. In the initial structure of the FMO protein a BChl a possessing a strained macrocycle was observed [25]. More recent structural studies on FMO proteins have also invoked changes in the planearity of their bound BChl a to account for their observed electronic properties[26]. Raman experiments, obtained in pre-resonance conditions with the Q$_y$ transition of this protein unambiguously show that the BChl responsible for the lower energy absorption transition is highly distorted (data not shown, Robert B in preparation). In the FMO protein, the coupling of these additional modes in FLN spectra thus likely arises from the distorted configuration of the terminal acceptor pigment in this protein.

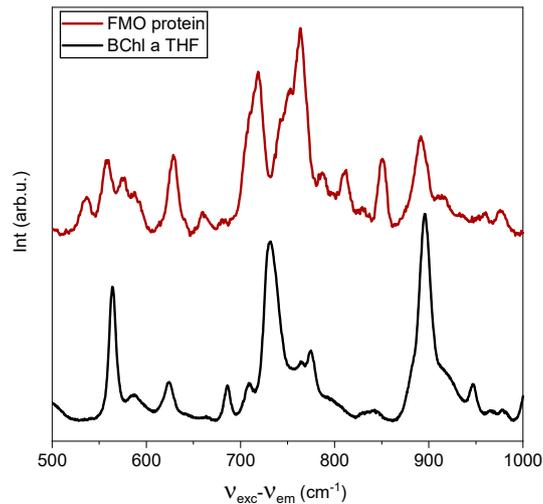

**Figure 3 |** Fluorescence line narrowing spectra (mid frequency region of FMO protein (red line, excitation wavelength 827 nm) and Bchl a (black line, excitation wavelength 822nm).

The vibronic landscape of the terminal emitters of structurally unrelated BChl-containing antenna (LH1, FMO) thus appears to generally contain unexpected additional modes coupled to their lower energy electronic transition, when compared to isolated molecules, modes which potentially open additional channels for excitation energy transfer in these proteins through vibrational assistance. This situation is quite different from the Chl-containing proteins from eukaryotic organisms. FLN spectra of LHCII, the major light-harvesting complex of photosystem 2 [27], of the minor antenna CP29 [28] and of photosystem 2 itself [29] have been published. In the FLN spectra of these Chl-containing protein, no band broadening is observed as compared to the FLN spectrum of isolated Chl a (figure 4).

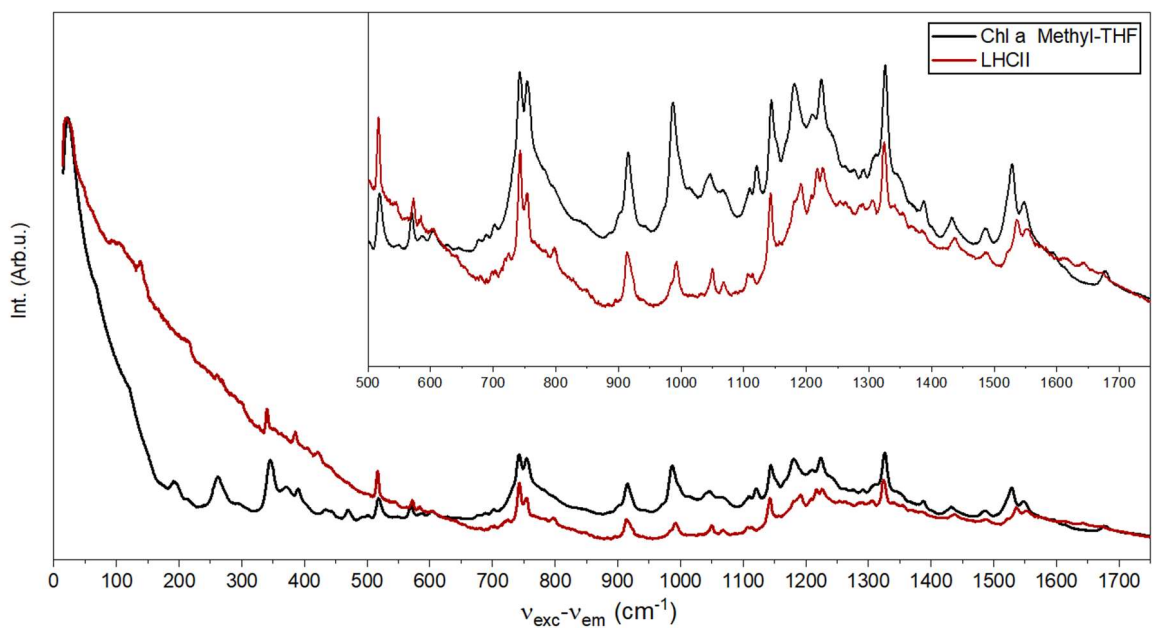

**Figure 4 |** Fluorescence line-narrowing spectra recorded at 4.2 K for isolated chlorophyll a in methyl-THF (bottom) and LHCII (top). Excitation wavelengths were 676.7nm (Chl a) and 682 nm (LHCII).

The bands arising from stretching mode of the conjugated keto C=O, the frequency of which is exquisitely sensitive to the molecular environment, are visible in FLN spectra of Chl a. In all the spectra, it is essentially constituted of a single component, indicating they contain the contribution of a single Chl molecule. In addition, FLN spectra published to date from Chl-containing proteins generally look extremely similar to those of isolated chlorophyll a (refs [27-29] and figure 4). The absence of new vibronic contributions in the spectra of protein-bound Chl molecules suggests that in Chl a-containing antenna vibrationally-assisted excitation energy transfers occur through vibrational modes of Chl a molecules in an equilibrium configuration, in contrast with what we have observed in BChl-containing light-harvesting complexes.

**Methods**

*Sample preparation.* Blastochloris viridis cells were grown photoheterotrophically in Böse's medium[30] at 28°C in flat sided glass medical bottles between banks of incandescent bulbs (Osram), harvested, and the RC-LHI membranes prepared essentially as described previously[31]. BChl b was extracted from *Blc. viridis* membranes using Tetrahydrofuran (THF)[14]. Integrity of the sample and BChl b purity relative to other pigments was controlled by electronic absorption spectroscopy.

Similarly, Chl a (Sigma-Aldrich) was re-purified in the presence of THF. The Fenna Matthews Olson (FMO) protein, isolated from *Chlorobium tepidum* cells, was prepared as described previously [32]. LHCII proteins were isolated from unstacked spinach chloroplasts as described[33].

***Fluorescence Line Narrowing (FLN) measurements***. A drop of concentrated sample was deposited on a glass slide (2-3 $\mu$L, OD >10) and frozen in liquid nitrogen prior to insertion in a helium-flow cryostat (SMC-TBT, Air Liquide, Sassenage, France). Excitation between 680-1000 nm range was provided by a Ti:sapphire laser (3900S, Spectra-Physics) pumped by a 10 W Sabre Innova laser (Coherent). The majority of the FLN spectra were recorded with 90° signal collection using a two-stage monochromator (U1000, Jobin-Yvon), equipped with 600 groove/mm gratings and a front-illuminated, deep-depleted CCD detector (Jobin-Yvon). Typically, less than 1 $\mu$W reached the sample, and the measurement time of individual spectra was one second. To avoid spurious hole-burning phenomena during the recording of FLN spectra, which would result in a decrease in the narrow vibrational bands after the first seconds of illumination, the samples were heated to 50 K between individual FLN measurements recorded at 4 K. FLN spectra of *Blc. viridis* RC-LH1 membranes were recorded at 4 cm$^{-1}$ resolution with back-scattering geometry using a Bruker IFS 66 infrared spectrophotometer coupled to a Bruker FRA 106 Raman module equipped with a continuous 1064 nm Nd:YAG laser, as described previously [34].

## Acknowledgements

Spectroscopic measurements were performed on the were performed on the Electronic and Vibrational Platforms of the I2BC, supported by the French Infrastructure for Integrated Structural Biology (FRISBI) ANR-10-INBS-05, the French National Agency for Research (EXCIT grant N°: ANR-20-CE11-0022, ELECTROPHYLL grant N°: ANR-21-CE50-0028-02), and the Infrastructures en Biologie Santé et Agronomie (IBiSA).

## Author contributions

Manuel J. Llansola-Portoles contributed to performing the spectroscopic measurements, purification of (B)Chls, and organization and conception of the manuscript.

James Sturgis participated to the conception of this work and contributed to performing the spectroscopic measurements

Andrew Gall contributed to preparing the proteins from bacteria and performing the spectroscopic measurements

Andrew Pascal contributed to preparing the proteins from oxygenic organisms

Leonas Valkunas contributed to interpreting the spectroscopic data

Bruno Robert designed the experiments, had the original idea, performed the analysis of the data, and wrote the initial draft of the manuscript.

## Competing interests

The authors declare not having any competing interests.

## Materials & Correspondence

The correspondence should be addressed to Bruno ROBERT.

E-mail: bruno.robert@cea.fr

## Extended Data

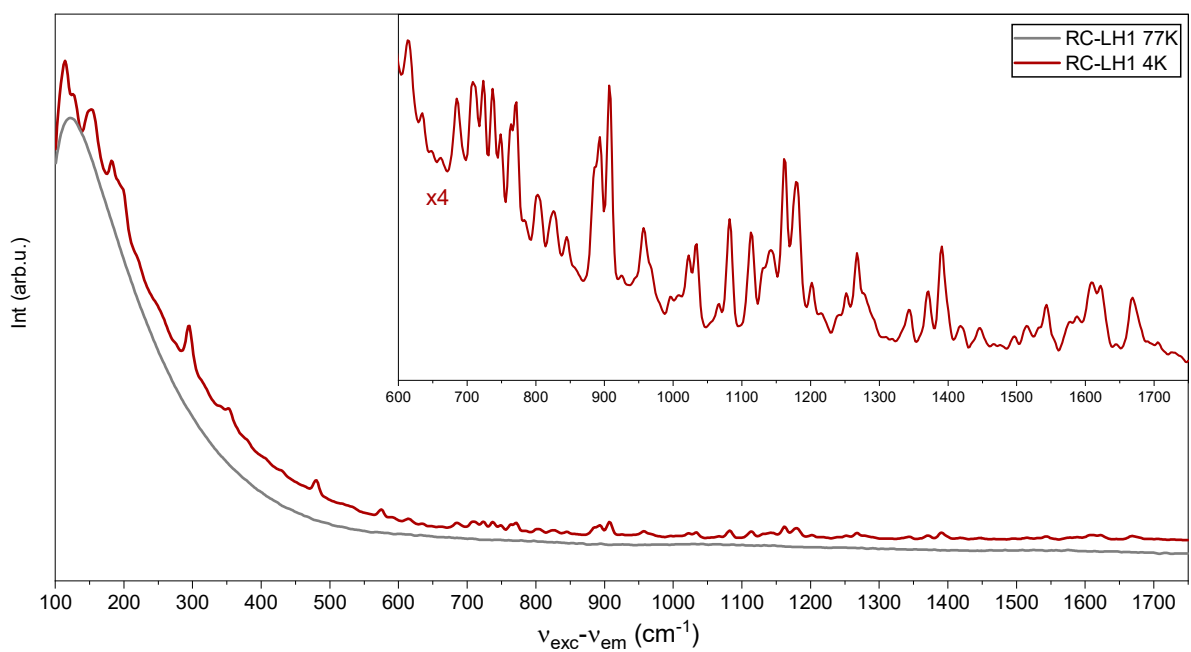

**Extended Data Fig. 1**: *Blc viridis* RC-LH1 emission (excitation 1064 nm, Red : 4K, Grey 77K) **Inset** : (x4) enhanced mid to high frequencies for clarity for RC-LH1.

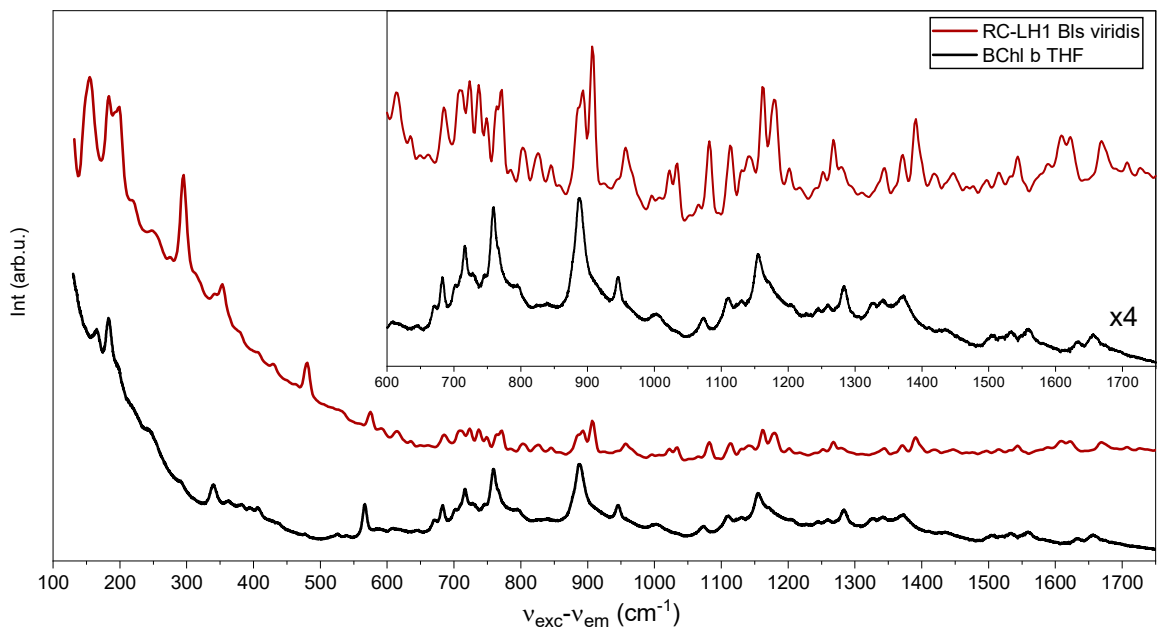

**Extended Data Fig. 2**: Full FLN spectra for BChl b (black line, excitation 820 nm) and RC-LH1 *Blc. viridis* (red line, excitation 1064 nm). **Inset**: Region between 600 and 1700 is shown after x4 enhancement for clarity

# References

1   Sauer, K. & Austin, L. A. Bacteriochlorophyll-protein complexes from the light-harvesting antenna of photosynthetic bacteria. *Biochemistry* **17**, 2011-2019, doi:10.1021/bi00603a033 (1978).

2   Trinkunas, G. *et al.* Exciton Band Structure in Bacterial Peripheral Light-Harvesting Complexes. *J. Phys. Chem. B* **116**, 5192-5198, doi:10.1021/jp302042w (2012).

3   van Grondelle, R., Dekker, J. P., Gillbro, T. & Sundstrom, V. Energy transfer and trapping in photosynthesis. *Biochim. Biophys. Acta, Bioenerg.* **1187**, 1-65, doi:10.1016/0005-2728(94)90166-X (1994).

4   Amerongen, H. v., Valkunas, L. & Grondelle, R. v. *Photosynthetic Excitons*. (World Scientific Publishing Co. Pte. Ltd., 2000).

5   Novoderezhkin, V., Marin, A. & van Grondelle, R. Intra- and inter-monomeric transfers in the light harvesting LHCII complex: the Redfield-Forster picture. *Phys. Chem. Chem. Phys.* **13**, 17093-17103, doi:10.1039/C1CP21079C (2011).

6   Sundström, V., Pullerits, T. & van Grondelle, R. Photosynthetic Light-Harvesting: Reconciling Dynamics and Structure of Purple Bacterial LH2 Reveals Function of Photosynthetic Unit. *J. Phys. Chem. B* **103**, 2327-2346, doi:10.1021/jp983722+ (1999).

7   McDermott, G. *et al.* Crystal structure of an integral membrane light-harvesting complex from photosynthetic bacteria. *Nature* **374**, 517-521 (1995).

8   Cogdell, R. J., Gall, A. & Köhler, J. The architecture and function of the light-harvesting apparatus of purple bacteria: from single molecules to in vivo membranes. *Q. Rev. Biophys.* **39**, 227-324, doi:10.1017/S0033583506004434 (2006).

9   Swainsbury, David J. K., Qian, P., Hitchcock, A. & Hunter, C. N. The structure and assembly of reaction centre-light-harvesting 1 complexes in photosynthetic bacteria. *Biosci. Rep.* **43**, doi:10.1042/bsr20220089 (2023).
10  Monshouwer, R., Abrahamsson, M., van Mourik, F. & van Grondelle, R. Superradiance and Exciton Delocalization in Bacterial Photosynthetic Light-Harvesting Systems. *J. Phys. Chem. B* **101**, 7241-7248, doi:10.1021/jp963377t (1997).
11  Monshouwer, R., Visschers, R. W., Mourik, F. v., Freiberg, A. & Grondelle, R. v. Low-temperature absorption and site-selected fluorescence of the light-harvesting antenna of Rhodopseudomonas viridis. Evidence for heterogeneity. *Biochim. Biophys. Acta, Bioenerg.* **1229**, 373-380, doi:10.1016/0005-2728(95)00020-J (1995).
12  Novoderezhkin, V., Monshouwer, R. & van Grondelle, R. Disordered Exciton Model for the Core Light-Harvesting Antenna of Rhodopseudomonas viridis. *Biophys. J.* **77**, 666-681, doi:10.1016/S0006-3495(99)76922-5 (1999).
13  Avarmaa, R. A. & Rebane, K. K. High-resolution optical spectra of chlorophyll molecules. *Spectrochimica Acta Part A: Molecular Spectroscopy* **41**, 1365-1380, doi:10.1016/0584-8539(85)80189-6 (1985).
14  Näveke, A. *et al.* Resonance Raman spectroscopy of metal-substituted bacteriochlorophylls: characterization of Raman bands sensitive to bacteriochlorin conformation. *J. Raman Spectrosc.* **28**, 599-604, doi:10.1002/(SICI)1097-4555(199708)28:8 (1997).
15  Robert, B. & Lutz, M. Structures of antenna complexes of several Rhodospirillales from their resonance Raman spectra. *Biochim. Biophys. Acta, Bioenerg.* **807**, 10-23, doi:10.1016/0005-2728(85)90048-9 (1985).
16  Koolhaas, M. H. C., van der Zwan, G. & van Grondelle, R. Local and Nonlocal Contributions to the Linear Spectroscopy of Light-Harvesting Antenna Systems. *J. Phys. Chem. B* **104**, 4489-4502, doi:10.1021/jp9918149 (2000).
17  Lapouge, K. *et al.* Conformation of Bacteriochlorophyll Molecules in Photosynthetic Proteins from Purple Bacteria. *Biochemistry* **38**, 11115-11121, doi:10.1021/bi990723z (1999).
18  Cogdell, R. J. *et al.* The structure and function of the LH2 (B800–850) complex from the purple photosynthetic bacterium Rhodopseudomonas acidophila strain 10050. *Prog. Biophys. Mol. Biol.* **68**, 1-27, doi:10.1016/S0079-6107(97)00010-2 (1997).
19  Trinkunas, G., Herek, J. L., Polívka, T., Sundström, V. & Pullerits, T. Exciton Delocalization Probed by Excitation Annihilation in the Light-Harvesting Antenna LH2. *Phys. Rev. Lett.* **86**, 4167-4170, doi:10.1103/PhysRevLett.86.4167 (2001).
20  Lutz, M., Kleo, J. & Reiss-Husson, F. Resonance raman scattering of bacteriochlorophyll, bacteriopheophytin and spheroidene in reaction centers of Rhodopseudomonas spheroides. *Biochem. Biophys. Res. Commun.* **69**, 711-717, doi:10.1016/0006-291X(76)90933-5 (1976).
21  Olson, J. M. The FMO Protein. *Photosynth. Res.* **80**, 181-187, doi:10.1023/B:PRES.0000030428.36950.43 (2004).
22  Rätsep, M., Blankenship, R. E. & Small, G. J. Energy Transfer and Spectral Dynamics of the Three Lowest Energy Qy-States of the Fenna-Matthews-Olson Antenna Complex. *J. Phys. Chem. B* **103**, 5736-5741, doi:10.1021/jp990918g (1999).
23  Wendling, M. *et al.* Electron−Vibrational Coupling in the Fenna−Matthews−Olson Complex of Prosthecochloris aestuarii Determined by Temperature-Dependent Absorption and Fluorescence Line-Narrowing Measurements. *J. Phys. Chem. B* **104**, 5825-5831, doi:10.1021/jp000077+ (2000).
24  Cheng, Y.-C & Fleming, G. R. Dynamics of Light Harvesting in Photosynthesis. *Annu. Rev. Phys. Chem.* **60**, 241-262, doi:10.1146/annurev.physchem.040808.090259 (2009).
25  Tronrud, D. E., Schmid, M. F. & Matthews, B. W. Structure and X-ray amino acid sequence of a bacteriochlorophyll a protein from Prosthecochloris aestuarii refined at 1.9 Å resolution. *J. Mol. Biol.* **188**, 443-454, doi:10.1016/0022-2836(86)90167-1 (1986).

26  Li, Y.-F., Zhou, W., Blankenship, R. E. & Allen, J. P. Crystal structure of the bacteriochlorophyll a protein from Chlorobium tepidum. *J. Mol. Biol.* **271**, 456-471, doi:10.1006/jmbi.1997.1189 (1997).
27  Peterman, E. J. G., Pullerits, T., van Grondelle, R. & van Amerongen, H. Electron–Phonon Coupling and Vibronic Fine Structure of Light-Harvesting Complex II of Green Plants: Temperature Dependent Absorption and High-Resolution Fluorescence Spectroscopy. *J. Phys. Chem. B* **101**, 4448-4457, doi:10.1021/jp962338e (1997).
28  Pascal, A. *et al.* Structure and Interactions of the Chlorophyll a Molecules in the Higher Plant Lhcb4 Antenna Protein. *J. Phys. Chem. B* **104**, 9317-9321, doi:10.1021/jp001504m (2000).
29  Peterman, E. J. G., van Amerongen, H., van Grondelle, R. & Dekker, J. P. The nature of the excited state of the reaction center of photosystem II of green plants: A high-resolution fluorescence spectroscopy study. *Proc. Natl. Acad. Sci.* **95**, 6128-6133, doi:10.1073/pnas.95.11.6128 (1998).
30  S.K., B. in *Bacterial Photosynthesis* (eds H. Guest, A. San Pietro, & L.P. Vernon) 501-519 (1963).
31  Evans, M. B., Hawthornthwaite, A. M. & Cogdell, R. J. Isolation and characterisation of the different B800–850 light-harvesting complexes from low- and high-light grown cells of Rhodopseudomonas palustris, strain 2.1.6. *Biochim. Biophys. Acta, Bioenerg.* **1016**, 71-76, doi:10.1016/0005-2728(90)90008-R (1990).
32  Francke, C., Otte, S. C. M., Miller, M., Amesz, J. & Olson, J. M. Energy transfer from carotenoid and FMO-protein in subcellular preparations from green sulfur bacteria. Spectroscopic characterization of an FMO-reaction center core complex at low temperature. *Photosynth. Res.* **50**, 71-77, doi:10.1007/BF00018222 (1996).
33  Ruban, A. V., Lee, P. J., Wentworth, M., Young, A. J. & Horton, P. Determination of the Stoichiometry and Strength of Binding of Xanthophylls to the Photosystem II Light Harvesting Complexes. *J. Biol. Chem.* **274**, 10458-10465, doi:10.1074/jbc.274.15.10458 (1999).
34  Mattioli, T. A. *et al.* Application of near-IR Fourier transform resonance Raman spectroscopy to the study of photosynthetic proteins. *Spectrochimica Acta Part A: Molecular Spectroscopy* **49**, 785-799, doi:10.1016/0584-8539(93)80103-H (1993).